\documentclass[12pt,english]{article}
\usepackage[margin=1in,footskip=0.25in]{geometry}

\usepackage{qcircuit}
\usepackage{tikz}
\usepackage{caption}
\usepackage{subcaption}
\usepackage{amsmath}
\usepackage{graphicx}
\usepackage{comment}
\usepackage{booktabs}

\graphicspath{{./figures/}}

\title{Constraint programming models for depth-optimal qubit assignment and SWAP-based routing} 

\author{Kyle E. C. Booth}

\date{%
\small Amazon Quantum Solutions Lab, Seattle, WA, USA, 98170 \\  \texttt{kybooth@amazon.com}}

\begin{document}

\maketitle

\begin{abstract}
Due to the limited connectivity of gate model quantum devices, logical quantum circuits must be compiled to target hardware before they can be executed. Often, this process involves the insertion of SWAP gates into the logical circuit, usually increasing the depth of the circuit, achieved by solving a so-called qubit assignment and routing problem. Recently, a number of integer linear programming (ILP) models have been proposed for solving the qubit assignment and routing problem to proven optimality. These models encode the objective function and constraints of the problem, and leverage the use of automated solver technology to find hardware-compliant quantum circuits. In this work, we propose constraint programming (CP) models for this problem and compare their performance against ILP for circuit depth minimization for both linear and two-dimensional grid lattice device topologies on a set of randomly generated instances. Our empirical analysis indicates that the proposed CP approaches outperform the ILP models both in terms of solution quality and runtime.
\end{abstract}

\section{Introduction}

The quantum circuit model of computation specifies quantum algorithms as sequences of logical quantum gates \cite{rieffel2011quantum}. Quantum circuit compilation is the process of compiling a logical quantum circuit to a target quantum device such that the compiled circuit adheres to device-specific connectivity constraints. Commonly, this process involves the insertion of SWAP gates into the original circuit enabling qubits to move to neighboring locations on the hardware. Determining the initial qubit allocation as well as when and where these gates should be inserted has been studied under the name \emph{qubit allocation and routing} \cite{cowtan2019qubit}. 

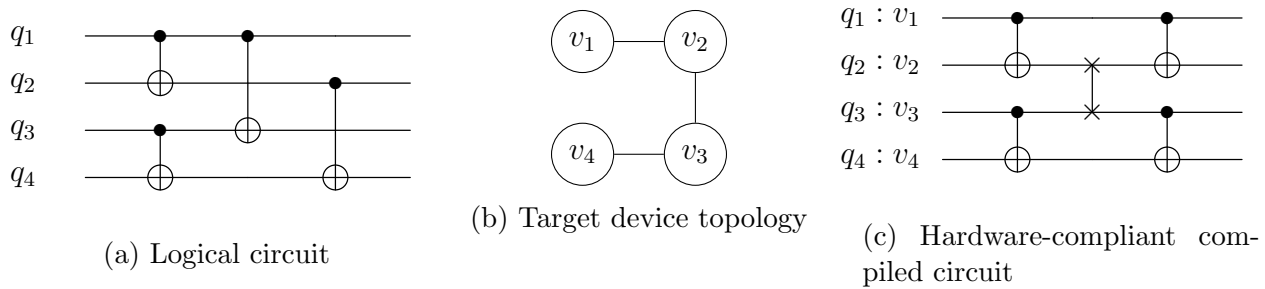
\begin{figure}[t!]

\begin{subfigure}[h]{0.32\textwidth}
\centering
\[
\Qcircuit @C=2.0em @R=0.7em @!R {
q_1 & & \ctrl{1} & \ctrl{2} & \qw & \qw\\
q_2 & & \targ    & \qw & \ctrl{2} & \qw\\
q_3 & & \ctrl{1} & \targ & \qw & \qw\\ 
q_4 & & \targ    & \qw & \targ & \qw 
}
\]
\caption{Logical circuit}
\label{example-logical-circuit}
\end{subfigure}
\hfill
\begin{subfigure}[h]{0.32\textwidth}
\centering
\begin{tikzpicture}
\node[draw, circle] (a) at (0, 0)  {$v_1$};
\node[draw, circle] (b) at (1.5, 0)  {$v_2$};
\node[draw, circle] (c) at (0, -1.5)  {$v_4$};
\node[draw, circle] (d) at (1.5, -1.5)  {$v_3$};
\draw[] (a) -- (b)  node[] {};
\draw[] (b) -- (d)  node[] {};
\draw[] (c) -- (d)  node[] {};
\end{tikzpicture}
\caption{Target device topology}
\label{example-target-device}
\end{subfigure}
\hfill
\begin{subfigure}[h]{0.32\textwidth}
\centering
\[
\Qcircuit @C=2.0em @R=0.7em @!R {
q_1: v_1 & & \ctrl{1} & \qw & \ctrl{1} & \qw \\
q_2: v_2 & & \targ    & \qswap & \targ  & \qw \\
q_3: v_3 & & \ctrl{1} & \qswap \qwx & \ctrl{1}  & \qw \\ 
q_4: v_4 & & \targ    & \qw & \targ  & \qw 
}
\]

\caption{Hardware-compliant compiled circuit}
\label{example-compiled-circuits}
\end{subfigure}
\caption{Qubit allocation and routing problem example specifying the input logical circuit (a), target device topology (b) and a compiled circuit (c). The input circuit has a depth of 2 (gates ($q_1,q_3$) and ($q_2,q_4$) can be executed in parallel), while the compiled circuit has a depth of 3.}
\label{fig:quantum-circuit-example}
\end{figure}

Figure \ref{fig:quantum-circuit-example} provides an illustration of an input logical quantum circuit, a quantum device topology (represented as an undirected graph that specifies device connectivity), and a valid compiled circuit that inserted a single SWAP gate. Due to the effects of quantum decoherence, particularly in superconducting devices, it is often beneficial to solve the qubit allocation and routing problem while minimizing circuit depth (where the depth is the number of layers of parallelized gates). For this reason, additional gates should be used sparingly and with high parallelization, when possible. 

A variety of techniques have been proposed for solving this problem in the literature, including both exact and heuristic approaches. Exact methods typically involve the use of search-based solvers leveraging smart inference techniques that, given enough time, will find and prove the optimal solution \cite{booth2018comparing, boccia2019SWAP, nannicini2021optimal, mulderij2020polynomial, venturelli2018compiling}. Alternatively, heuristic methods (which have, until recently, been the focus of previous work) sacrifice completeness in favor of rapidly producing high quality circuits \cite{li2019tackling,zulehner2018efficient,cowtan2019qubit}; these methods tend to scale more effectively to larger problem instances as well.

In this short paper, we focus on the former with the goal of developing exact model-based approaches that outperform those from the literature. Specifically, we introduce and report initial results for two new CP models for the qubit assignment and routing problem. Our models have a relatively simple implementation, and leverage constraints supported by a variety of CP solvers. The first model is designed for linear array quantum device topologies, while the second can be used to solve problems involving more general architectures. We conduct an empirical analysis of the ILP models proposed by Boccia et al. \cite{boccia2019SWAP} and Nannicini et al. \cite{nannicini2021optimal} using circuit depth minimization as our objective function.
We demonstrate, through empirical evaluation using different solvers, that our CP models outperform these existing ILP approaches in terms of solution quality and runtime for depth minimization. 

The remainder of this paper is organized as follows. In Section \ref{sec:problem-definition} we define the specific variant of the qubit assignment and routing problem that we consider in this paper. In Section \ref{sec:models} we present our new CP models for the studied problem. In Section \ref{sec:experiments} we conduct an empirical assessment of the presented models on both linear array and 2D lattice architectures of varying size. Finally, in Section \ref{sec:conclusions} we provide concluding remarks.

\section{Preliminaries and definitions}\label{sec:problem-definition}

We follow previously used notation with minor alterations to ease the presentation of the problem definition \cite{nannicini2021optimal}. The input to the qubit mapping and routing problem consists of: i) a hardware graph, and ii) a sequence of quantum gate groups. The hardware graph, $H=(V,E)$, consists of nodes $i \in V$, where each node represents a physical qubit on the quantum computer. The graph has edges $e \in E$, where each edge $e = \{i,j\}$ dictates pairs of physical qubits that can execute two-qubit gates (i.e., qubits that are neighboring eachother on the architecture). Often, as in previous work \cite{nannicini2021optimal}, it is useful to define a directed graph $A$ such that each undirected edge in $E$ corresponds to two directed edges in $A$, $(i,j)$ and $(j,i)$. In this paper we study both linear and general hardware graphs (e.g., lattices). 

The sequence of quantum gate groups is represented by $G = (G^1, G^2, \dots, G^L)$ where $G^{\ell}$ indicates the set of gates that must be executed in the $\ell^{th}$ layer of the logical circuit. In this paper, we consider two-qubit quantum gates, $g = \{p,q\}$, involving a pair of logical qubits $p,q \in Q$ such that $G^{\ell} = \{g^{\ell}_1, g^{\ell}_2, \dots \}$. All of the two-qubit gates executed in a given layer involve a set of non-overlapping pairs of qubits. This way, the gates in a layer are parallelized. To account for inserted SWAP gates, we augment the original circuit with a number of `dummy layers' between each of the logical circuit layers. We denote the total set of layers as $L'$, and use $L_{SWAP}$ to specify the dummy layers only.

Qubit \emph{assignment} involves producing an initial mapping of logical qubits, $Q$, to physical qubits, $V$, on the quantum hardware. For simplicity, we refer to logical qubits and physical qubits as qubits and nodes from now on, respectively. We also assume that $|Q| = |V|$.\footnote{Note that, in the case where $|Q| < |V|$, we can just introduce and route dummy qubits.} 

Qubit \emph{routing} is the process of moving qubits around the hardware architecture such that the gates at each layer can be executed in a way that satisfies the connectivity of the device. In this work, we consider routing accomplished via the use of SWAP gates (although we note there are other routing techniques \cite{bapat2021quantum}), two-qubit gates which exchange the position of two neighboring qubits. These SWAP gates are inserted into the dummy layers between the layers of the idealized quantum circuit and (often) increase its overall depth.

The objective of the qubit assignment and routing problem studied in this paper is to: i) assign qubits to nodes on the hardware, and ii) route the positions of the qubits such that the gates at each layer can be executed while satisfying the connectivity of the device. We seek to minimize the number of dummy layers utilized, effectively resulting in circuits with lower depth.  This problem is known to be NP-Complete \cite{botea2018complexity}.

Following previous work, our assumptions on the formulation of the problem include: 
\begin{itemize}
    \item Qubits involved in a two-qubit gate can also swap positions. This is also called `merging' SWAPs with adjacent two qubit gates \cite{jurcevic2021demonstration}.
    \item Two-qubit gates are undirectional (i.e., the cost of executing a two-qubit gate in the forward configuration is the same as in the reverse configuration). 
    \item We restrict our attention to circuits involving two-qubit gates, since single qubit gates can be merged with two-qubit gates and thus do not need to be considered. 
\end{itemize}

\section{Constraint programming models}\label{sec:models}

For each of our proposed models, the main decision variable is $x_{q\ell} \in \{1,\dots,|V|\}$ which represents the node location of logical qubit $q\in Q$ at layer $\ell \in L'$. Since ILP is less flexible than CP (e.g., due to linearity restrictions) the ILP models from the literature \cite{nannicini2021optimal,boccia2019SWAP,mulderij2020polynomial} must introduce additional variables to properly model the problem. In our models, it is sufficient to constrain the values of $x_{q\ell}$ from one layer to the next.

\subsection{Linear array architectures}

The first CP model we present is applicable to linear array architectures (such as the one presented in Figure \ref{example-target-device}). This model is motivated by previous work in ILP \cite{mulderij2020polynomial},\footnote{We note that the work in \cite{mulderij2020polynomial} does not minimize circuit depth, but rather the number of SWAPs inserted, and so we do not compare to it in this paper (non-trivial changes to the model must be made in order to express a depth minimization objective).} and employs absolute value constraints to ensure that qubit movement between layers is valid. 

The first constraint in our model ensures that, at each layer, each logical qubit is located at one node on the architecture, and all of the locations of the logical qubits are different. We use the $\texttt{alldifferent}$ global constraint \cite{van2001alldifferent} to accomplish this: 
\begin{equation}
    \texttt{alldifferent}(\{x_{1\ell}, \dots, x_{|Q|\ell}\}), \forall \ell \in L' \label{linear:alldiff}
\end{equation}
The next set of constraints ensure that the gates specified in each layer are executed while adhering to the connectivity constraints of the target hardware. These constraints are expressed as follows:
\begin{equation}
    |x_{p\ell} - x_{q\ell}| = 1, \forall \ell \in L, g = (p,q) \in G^\ell
\end{equation}
Intuitively, these constraints specify that the locations of logical qubits $(p,q)$ involved in two-qubit gate $g$ must be neighboring. Note that, since this model is for a linear architecture, this constraint is enough to ensure the gates in a layer can be executed; more complex architectures require more sophisticated modeling (as described in the next section). 

A similar set of constraints is used to constrain the movement of qubits from one layer to the next. These are expressed as follows:
\begin{equation}
|x_{q\ell} - x_{q\ell+1}| \leq 1, \forall \ell \in \{1,\dots, |L'|-1\}, q \in Q 
\end{equation}
The above constraints dictate that, from one layer to the next, a given qubit cannot move more than one location away from its current location. An additional set of constraints must be added in addition to these to ensure that pairs of qubits, such that one qubit is involved in a gate at a given layer and the other is not (recall that the qubit pairs involved in a gate are permitted to SWAP), cannot SWAP from that layer to the next. This constraint is expressed as follows:
\begin{equation}
  |x_{p\ell} - x_{q\ell+1}| \leq 1  , \forall \ell \in L, g = (p,q) \in G^\ell \label{linear:restrict}
\end{equation}
Constraint \eqref{linear:restrict} ensures that the qubit pair involved in a gate are still neighboring at the next layer. Note that this permits the qubit pair to SWAP places, but it does not permit either qubit swapping with another qubit not involved in the gate. 

Finally, we encode the objective function of the optimization which is to minimize the number of dummy layers added to the circuit to support SWAP insertions; this is effectively the same as circuit depth, and follows previous work \cite{nannicini2021optimal}. To express this objective, we introduce a binary decision variable $z_\ell \in \{0,1\}$ for each layer, $\ell \in L_{SWAP}$. We constrain the variable such that it takes on a value of $1$ whenever a SWAP gate is inserted into a dummy layer:
\begin{equation}
z_\ell \geq |x_{q\ell} - x_{q\ell+1}|, \forall q \in Q, \ell \in L_{SWAP}. \label{linear:objective-cons}
\end{equation}
Constraint \eqref{linear:objective-cons} tracks each time a qubit changes locations from one layer to the next, and the location change is not due to a (non-SWAP) gate operation. Finally, the objective function is expressed as:
\begin{equation}
    \min \sum_{\ell \in L_{SWAP}} z_\ell
\end{equation}
Note that if all qubit routing operations can be achieved by merged SWAPs, the optimal solution will be zero as no auxiliary SWAP gates need to be introduced. 

The linear architecture CP model has $O(|Q||L'|)$ variables, each with a domain of $O(|V|)$, and 
$O(|Q||L'|)$ constraints. While the model is designed according to the assumptions stated in Section \ref{sec:problem-definition}, it can be altered fairly easily to accommodate problem variations. For example, if qubits involved in a two-qubit gate \emph{cannot} also SWAP places, we could simply constrain the locations of the qubits involved in the gate as necessary; though this would undoubtedly increase the depth of the produced circuits. 

\subsection{General architectures}

When the problem is extended to more general architectures, the linear architecture model is no longer valid. This is because in the linear model, there is a clear mapping between the locations of the logical qubits and the chip connectivity; any qubits whose positions are within one of eachother could have a gate applied between them. In the more general case, however, this is no longer true.\footnote{In a simple square toplogy with four nodes, $v_1$ and $v_3$ can be neighboring, but their node labels are not one apart.}

Our model for general architectures is similar to that for linear array architectures, however, instead of using absolute value constraints we use \texttt{table} constraints. The \texttt{table} constraint specifies the list of tuples (solutions) to which a vector of variables can be fixed. For example, the constraint \texttt{table}$((y_1, y_2), \{(1,2), (2,3)\})$ specifies that, in a solution, the variables $y_1$ and $y_2$ can be assigned values $(1,2)$ or $(2,3)$, respectively. 

For our model, we utilize the \texttt{table} constraint to conveniently encode the connectivity of the hardware device. Our model for general architectures is given as follows:
\begin{align}
    \min \quad  & \sum_{\ell \in L_{SWAP}} z_\ell & \\
    & \texttt{alldifferent}(\{x_{1\ell},\dots, x_{|Q|\ell}\}) & \forall \ell \in L' \label{gen:alldiff} \\ 
    & \texttt{table}((x_{p\ell}, x_{q\ell}), A) & \forall \ell \in L, g = (p,q) \in G^\ell \label{gen:gates} \\
     & \texttt{table}((x_{q\ell}, x_{q\ell+1}), A \cup \{(i,i) : i \in V\}) & \forall q \in Q, \forall \ell \in \{1,\dots,|L'|-1\} \label{gen:SWAP} \\
     & x_{q\ell} = x_{q\ell+1} \vee x_{q\ell} = x_{p\ell+1} & \forall \ell \in L, g = (p,q) \in G^\ell \label{gen:restrict} \\
     & z_\ell|Q| \geq |x_{q\ell} - x_{q\ell+1}| & \forall \ell \in L_{SWAP}, q \in Q  \label{gen:obj}
\end{align}

The first constraint in the general architecture model, Constraint \eqref{gen:alldiff}, is the same as Constraint \eqref{linear:alldiff} in the linear array model, leveraging the \texttt{alldifferent} global constraint. Constraint \eqref{gen:gates} uses the \texttt{table} constraint to ensure that, at each gate layer, the qubits involved in each gate are at neighboring locations on the architecture (i.e., at locations represented by one of the arcs in arcset $A$). Constraint \eqref{gen:SWAP} uses $\texttt{table}$ to  dictate the permissible movement of qubits from one layer to the next, and Constraint \eqref{gen:restrict} adds some additional restriction regarding the movement of qubits involved in a two-qubit gate. Finally, Constraint \eqref{gen:obj} links the objective function to the main variables, acting as a flag each time a qubit changes positions from one layer to the next (not including original circuit layers). 

As with the linear array model, the general architecture model also has $O(|Q||L'|)$ variables and constraints.

\section{Experiments}\label{sec:experiments}

We present an empirical analysis of our models on a benchmark set of randomly generated instances. Following previous work \cite{nannicini2021optimal} we generate `square' circuits (similar to those used for quantum volume testing) where $|Q|=|L|$ (e.g., $4\times 4$ indicates a circuit with four qubits and four layers). For a given problem size, we generate a random permutation of the qubits $\{1,2\dots,|Q|\}$ for each $\ell \in L$. Based on this permutation, we group neighboring pairs of qubits into the $\lfloor\frac{|Q|}{2}\rfloor$ two-qubit gates for that layer (e.g., permutation 3-1-2-4 would yield gates $(3,1)$ and $(2,4)$). By increasing the size of $Q$ and $L$, we test the scalability of our models (e.g., a $10 \times 10$ instance will involve an initial circuit with 50 two-qubit gates). For each problem size, we generate ten problem instances. 

The target devices used for experimentation include a linear array and two-dimensional lattice grids of increasing size. We generate the devices such that the number of hardware vertices, $|V|$, is equal to the number of logical qubits, $|Q|$.\footnote{Our approaches are not restricted to this case; dummy qubits can be used to handle the situation where $|V| > |Q|$.} The linear array device topologies are straightforward. The lattice grids are generated for $|Q| \in \{4, 6, 8, 9, 10\}$, where $Q\in \{4,9\}$ corresponds to square lattices, and the remainder are rectangular lattices.

The CP models are implemented with the CP-SAT solver in OR-Tools (v9.3) \cite{perron2011operations} using the Python interface. We include the symmetry breaking constraints used in previous ILP models \cite{nannicini2021optimal}. The absolute value expressions in the CP models were implemented using auxiliary variables and the \texttt{AddAbsEquality} constraint, while the \texttt{table} constraint was implemented using the \texttt{AddAllowedAssignments} constraint. Finally, in the general architecture CP model, the dijsunction in Constraint \eqref{gen:restrict} was implemented using auxiliary boolean variables and the \texttt{OnlyEnforceIf} enforcement literal. Following previous work \cite{nannicini2021optimal}, we set the number of dummy layers between each original layer to four. 

To thoroughly assess the previously proposed ILP models, we implement them in SCIP \cite{achterberg2009scip}, Xpress (v8.13.5), and using the CP-SAT solver in OR-Tools.\footnote{Since the ILP models do not contain continuous variables, this is straightforward.} For the SCIP experiments, we use the OR-Tools modeling interface and select SCIP as the backend solver. All experiments are run with default search and inference settings on a machine with a 2.6 GHz 6-Core Intel i7 processor and 16GB of RAM. Additionally, some modifications to the model presented by Boccia et al. are made to enable fair comparison. Specifically, auxiliary variables and constraints were added to permit the merged SWAP functionality.

\begin{figure*}[t]
    \centering
    \begin{subfigure}[t]{0.5\textwidth}
        \centering
        \includegraphics[height=2.5in]{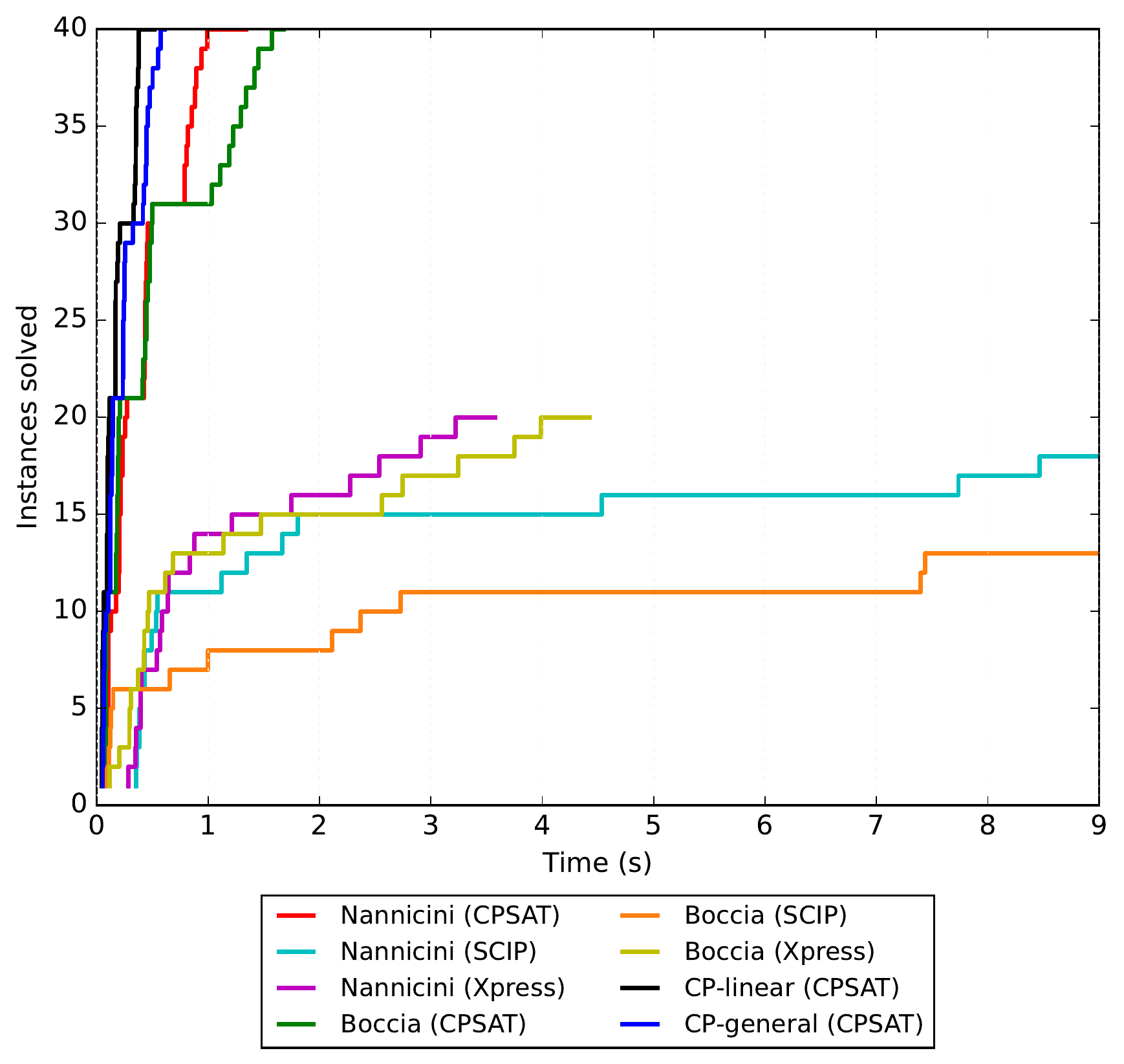}
        \caption{Class 1 instances ($|Q| \in \{4,5,6,7\}$)}
        \label{fig:linear-results-small}
    \end{subfigure}%
    ~ 
    \begin{subfigure}[t]{0.5\textwidth}
        \centering
        \includegraphics[height=2.5in]{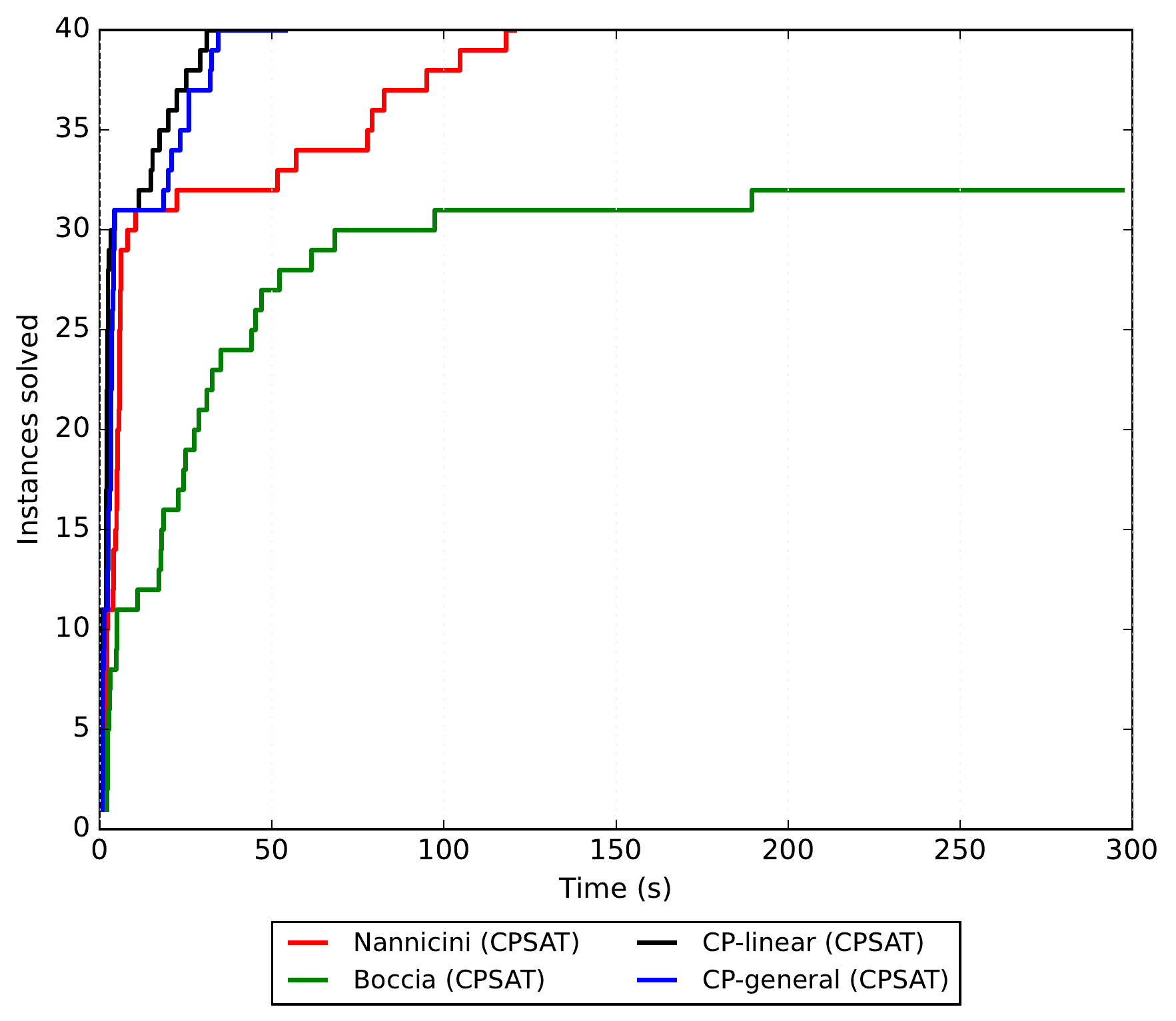}
        \caption{Class 2 instances ($|Q| \in \{8,9,10,11\}$)}
        \label{fig:linear-results-medium}
    \end{subfigure}
    \caption{Empirical results: CP models against ILP models from the literature for linear array device topologies. Number of instances solved to proven optimality over time. Time limit of 10 seconds (Class 1) and five minutes (Class 2).}
    \label{fig:linear-results}
\end{figure*}

\subsection{Linear array architectures} 
Our first set of experiments involve linear array architectures. For these experiments, both of the CP models we propose can be used to solve the problem. The results are visualized in Figure \ref{fig:linear-results}.\footnote{Our proposed models are denoted `CP-linear' (for linear array model) and `CP-general' (for general model).} We split the instances into two classes: the first involves randomly generated square circuits of size $ 4 \times 4$, $ 5 \times 5$, $6 \times 6$, and $7 \times 7$ while the second involves square circuits of sizes $8 \times 8$, $9 \times 9$, $10 \times 10$, and $11 \times 11$. 

In Figure \ref{fig:linear-results-small}, we illustrate the performance of all of the implementations on Class 1 instances with a solver time limit set to 10 seconds. The figure details the number of instances solved to proven optimality vs. runtime. We can see that all of the methods implemented with the CP-SAT solver in OR-Tools are able to quickly find and prove the optimal solution to all 30 problems in less than two seconds. Conversely, the ILP models implemented in SCIP and Xpress are much slower at finding and proving optimal solutions. For these approaches, the SCIP implementations seem to perform the worst, and Xpress provides a modest improvement. The experiments indicate that ILP (and ILP-based solvers) may not be the best candidate approach for this problem; this is in-line with previously reported results that showed ILP (implemented with an ILP solver) struggled to solve square circuit instances to proven optimality in short runtimes beyond six qubits \cite{nannicini2021optimal}. As such, we elect to only investigate the models solved with OR-Tools CP-SAT for larger problem instances.

Figure \ref{fig:linear-results-medium} illustrates the performance of the CP-based implementations on medium-sized instances with a solver time limit of 5 minutes. From the figure, it is evident that our proposed CP models outperform the ILP models even when using the OR-Tools CP-SAT solver for all methods. Further, the linear array model (using absolute value constraints) performs slightly better than the more general model. Both models are able to find and prove optimality for all problem instances in less than 50 seconds, whereas the implementations of the ILP methods require significantly more time. 

Recall that, due to the insertion of SWAP gates as a result of our optimization, the number of layers in the final compiled circuits almost always increases. In Table \ref{table:linear-depths} we detail the average circuit depth for each of the problem sizes, obtained by our depth-optimal CP methods. From the table, we can see that the inclusion of SWAP gates often doubles the depth of the circuit. 
 
\begin{table}[]
\centering
\begin{tabular}{cccc} \toprule
\textbf{Class 1} & \textbf{Compiled circuit depth (avg.)} & \textbf{Class 2} & \textbf{Compiled circuit depth (avg.)} \\ \midrule
$4 \times 4$ &  6.1 & $8 \times 8$   & 18.0 \\
$5 \times 5$ &  6.2 & $9 \times 9$   &  20.9 \\
$6 \times 6$ & 10.9 & $10 \times 10$   &  27.0 \\
      $7 \times 7$      &  12.4    & $11 \times 11$ & 30.7 \\ \bottomrule
\end{tabular}
\caption{Empirical results: CP model solution circuit depths by class, linear array device topologies.}
\label{table:linear-depths}
\end{table}

\begin{table}[b]
\centering
\begin{tabular}{cccc} \toprule
\textbf{Class 1} & \textbf{Compiled circuit depth (avg.)} & \textbf{Class 2} & \textbf{Compiled circuit depth (avg.)} \\ \midrule
$4 \times 4$ & 4.0 & $8 \times 8$   & 10.3 \\
$6 \times 6$ & 6.5 & $9 \times 9$   & 12.2 \\
    $-$         &   $-$  & $10 \times 10$ & 15.4 \\ \bottomrule
\end{tabular}
\caption{Empirical results: CP model solution circuit depths by class, 2D grid lattice device topologies.}
\label{table:lattice-depths}
\end{table}

\subsection{General architectures} 
\begin{figure*}[t]
    \centering
    \begin{subfigure}[t]{0.5\textwidth}
        \centering
        \includegraphics[height=2.5in]{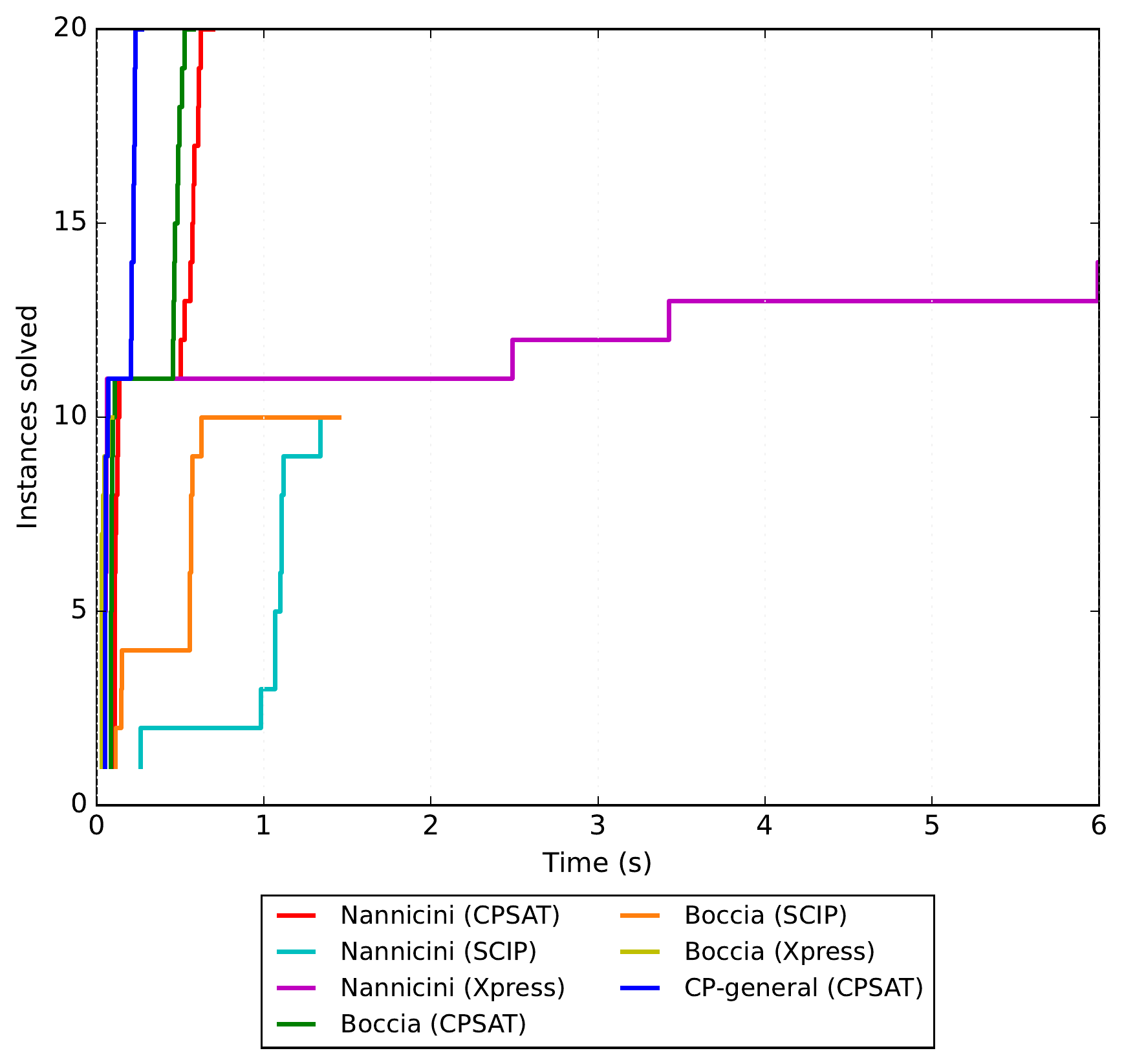}
        \caption{Class 1 instances ($4\times4$, $6\times6$)}
        \label{fig:lattice-results-small}
    \end{subfigure}%
    ~ 
    \begin{subfigure}[t]{0.5\textwidth}
        \centering
        \includegraphics[height=2.5in]{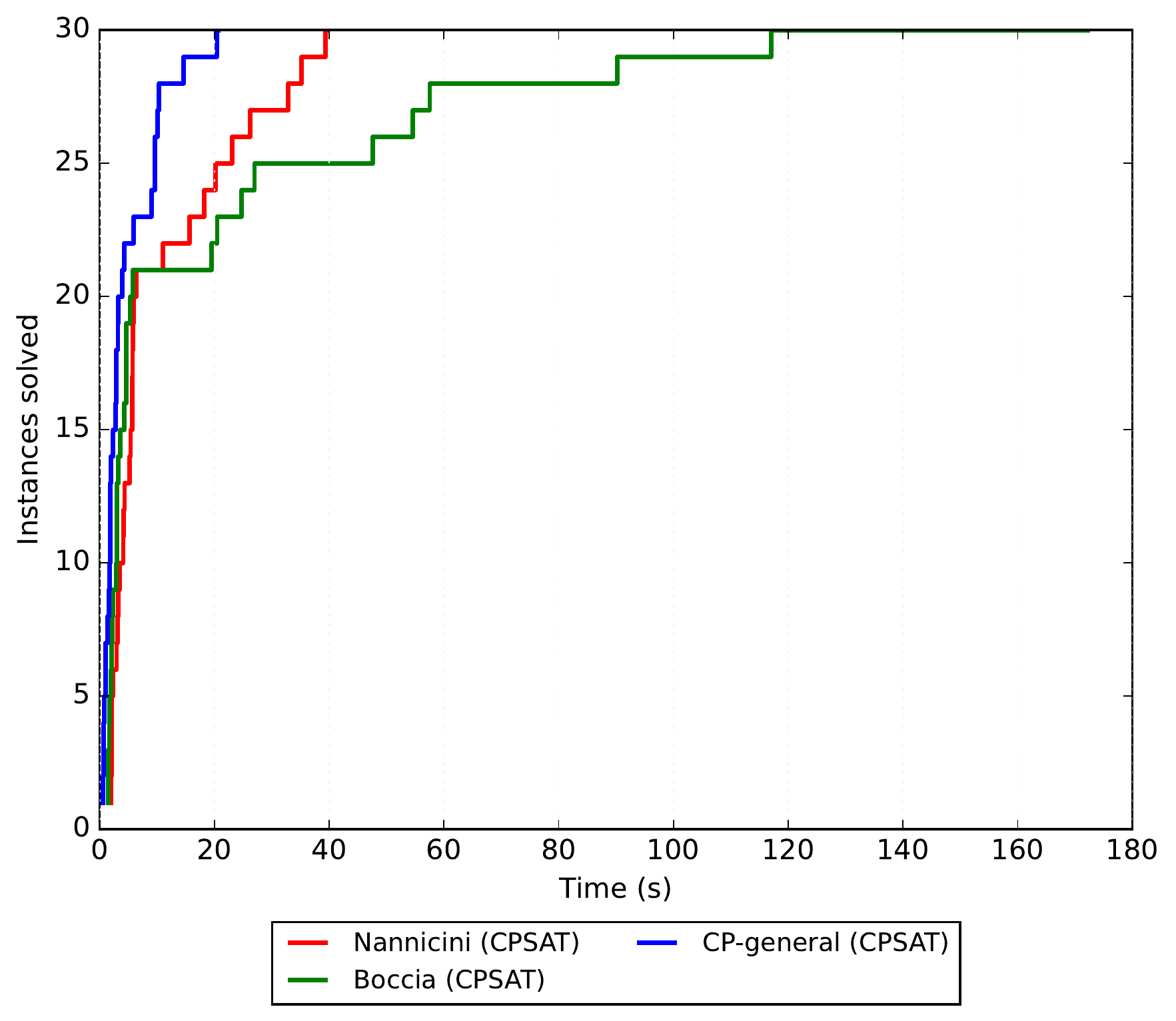}
        \caption{Class 2 instances ($8\times8$, $9\times9$, $10\times10$)}
        \label{fig:lattice-results-medium}
    \end{subfigure}
    \caption{Empirical results: CP models against ILP models from the literature for 2D grid lattice device topologies. Number of instances solved to proven optimality over time. Time limit of 10 seconds (Class 1) and five minutes (Class 2).}
    \label{fig:lattice-results}
\end{figure*}

Our second set of experiments involve 2D grid lattice topologies, and are visualized in Figure \ref{fig:lattice-results}. We use the same solvers and settings as in the linear array experiments (recall that we cannot run the proposed linear array model for these lattice topology problems). For these tests, Class 1 instances are $4 \times 4$ and $6 \times 6$ (to permit grid lattice construction), while Class 2 instances are $8 \times 8$, $9 \times9$, and $10 \times 10$.

In Table \ref{table:lattice-depths} we summarize the average depths of the optimally compiled circuits for each of the problem sizes. Comparing to the linear array results in Table \ref{table:lattice-depths}, it is immediately apparent that the increased connectivity offered by 2D lattice topologies results in significantly shorter circuits. In the $10 \times 10$ case, for example, only 5.4 layers were added (on average) to permit SWAP operations, versus the 17.0 layers added (on average) in the linear case.

In terms of model/solver runtime performance, from Figure \ref{fig:lattice-results-small} we see a similar trend to the linear topology case: the ILP methods implemented in SCIP and Xpress struggle to find and prove optimal solutions within the runtime limit, while the OR-Tools implementations rapidly solve these problems (in less than one second). Figure \ref{fig:lattice-results-medium} illustrates the performance of the OR-Tools implementations for the larger class of instances. As visualized in the figure, our proposed CP approach is able to find provably optimal solutions to all instances significantly faster than the ILP models from the literature.

\section{Conclusions}\label{sec:conclusions}

In this paper we propose CP models for depth-optimal qubit assignment and SWAP-based routing. Our first model is specific to linear array topologies, while our second model is applicable to more general architectures (e.g., grid lattices). We conduct a series of experiments on randomly generated circuits, and demonstrate that the CP-based approaches provide superior performance over their ILP counterparts. Our results suggest that CP is a promising technology for producing provably depth-optimal circuits when qubit routing is accomplished via SWAP gate insertion.

\bibliography{main}

\end{document}